# A First-principles study on $ABBr_3$ (A = Cs, Rb, K, Na; B = Ge, Sn) halide perovskites for photovoltaic applications

Dibyajyoti Saikia[1], Mahfooz Alam[2], Jayanta Bera[1], Atanu Betal[1], Appala Naidu Gandhi[2], Satyajit Sahu[1]

[1]Department of Physics, Indian Institute of Technology Jodhpur, Jodhpur, India 342037

[2]Department of Metallurgical and Materials Engineering, Indian Institute of Technology Jodhpur, Jodhpur, India 342037

**Abstract:** In recent years, halide perovskite-based solar cells have received intensive attention, and demonstrated power conversion efficiency as high as 25.8%. With regard to the toxicity of Pb and the instability of organic elements, all inorganic lead-free perovskites (ILPs) have been extensively studied to achieve comparable or greater photovoltaic performance. In order to develop ILPs as an alternative for solar cell applications, we performed first-principles calculations of $ABBr_3$ perovskites (A = Cs, Rb, K, and Na, and B = Sn, and Ge). Structural, electronic, and optical properties were systematically studied to probe the potentiality in photovoltaic applications. All these ILPs exhibited a direct bandgap in the range of 1.10 – 1.97 eV, highly beneficial for absorbing solar energy. Furthermore, these ILPs demonstrated significant optical absorption (over $10^5$ $cm^{-1}$) in the whole UV-Vis spectrum. These results will be helpful for designing highly efficient lead-free perovskite solar cells.

**Introduction:** Organic-inorganic halide perovskites (OIHPs) have gained significant interest as potential candidates in the field of photovoltaics industry due to ease of fabrication and optoelectronic properties [1]–[4]. Within a few years, the power conversion efficiency (PCE) of perovskite solar cells (PSCs) have reached certified values of over 25% [5], [6]. However, the organic elements (MA and FA) present in the OIHPs showed inferior instability under several environmental conditions [7]–[10],limiting the application. Replacement of these organic species with inorganic alkali cations ($Cs^+$ and $Rb^+$) overcome this problem [11]. All inorganic $CsPbX_3$ perovskites are considered as a potential candidate for efficient PSCs due to their high thermal stability and improved optoelectronic properties [12]. However, the toxicity that arises from the heavy toxic element Pb causes serious concern to the environment and constraints the large-scale commercialization [13], [14].

Replacement of $Pb^{2+}$ cation with other less toxic element could be considered as an efficient tool for the practical implementation of PSCs. Recently, substitution of Pb with other divalent cations like $Sn^{2+}$, $Ge^{2+}$, $Zn^{2+}$, $Cu^{2+}$, $Mn^{2+}$, etc., and other elements such as Bi, Sb, Ag, etc. have been widely investigated to explore their potentiality in photovoltaics as well as optoelectronics [15]–[19]. Among them, Sn and Ge are accounted to be key Pb replacements due to similar electronic characteristics [18]. Chen et al. reported 4.92% PCE on lead-free PSCs by synthesizing $CsGeX_3$ perovskite quantum rods [20], while mixed tin-germanium ($CsSn_{0.5}Ge_{0.5}I_3$) based PSCs delivered 7.11% PCE [19]. Meanwhile, $CsSnI_3$ based PSCs achieved over 10% PCE [21]. It was demonstrated that perovskite materials ($ABX_3$) with Br in the X-site position have shown superior stability compared to its iodine counterpart [22]. Upon substitution of I with Br, orthorhombic phase $CsSnI_3$ changes to cubic phase $CsSnBr_3$. In addition, Br addition improves the $V_{oc}$ of mixed $CsSnI_{3-x}Br_x$ PSCs [23]. In another work, Song et al. developed Sn-based PSCs in a reducing vapor atmosphere. The fabricated $CsSnBr_3$ based device delivered 3.04% PCE, while for $CsSnI_3$ based device PCE was 1.83% [24]. However, the PCE of all inorganic lead-free PSCs are still quite lower with regard to state-of-the-art OIHP-based PSCs. It could be expected that replacement of Cs with Rb in the A-site position improve the quality of perovskite film and thus, suitability in PSCs [25], [26]. However, to our knowledge, except a few theoretical investigations, no experimental work has been performed on $RbBBr_3$ (B = Sn, Ge), although there are several experimental and theoretical investigations on $CsBBr_3$ perovskites. Thus, a great deal of insight is required for the further development of ILPs.

Here, an investigation on $ABBr_3$ (A = Cs, Rb, K, Na; B = Ge, Sn) perovskites has been performed using first-principles calculations to explore their potential as photovoltaic materials. We have systematically studied the structural, electronic, and optical properties. Moreover, the photovoltaic performance of $ABBr_3$ halide perovskites has been explored.

## Computational Details:

First-principles calculations were carried out using the pseudopotential-based density functional theory as implemented in the Quantum Espresso package. Ultrasoft pseudopotentials were used for electron-ion interactions within the generalized gradient approximation using Perdew-Burke-Ernzerhof (PBE) functional. The electronic wave function cut-off and kinetic energy cut-off were set to 65 Ry and 520 Ry, respectively. For the structural relaxation, 10 × 10 × 10 k-grid was employed, while for the electronic structure calculation a denser k-grid of 15 × 15 × 15 was implemented. All the structures were fully relaxed until the residual force and energy on each atomic site was less than 0.01 eV/Å and $10^{-4}$ eV, respectively. To obtain a more accurate band structure, electronic structure calculations were carried out by employing hybrid functional (HSE06) as implemented in the VASP package. Furthermore, optical properties were calculated using the HSE06 functional.

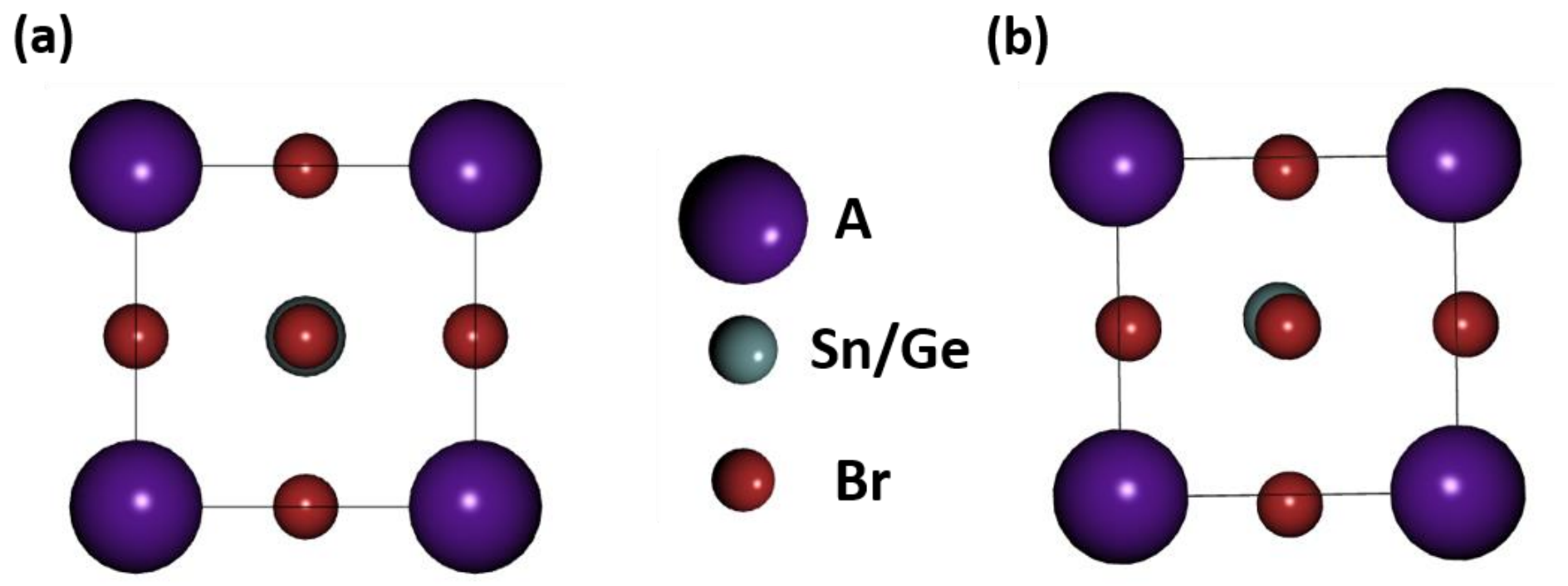

Figure 1. Crystal structure of **(a)** $ASnBr_3$ and **(b)** $AGeBr_3$ halide perovskites.

**Result and Discussion:**

**Structural properties:** At room temperature $CsSnBr_3$ exhibits a cubic phase with the space group of *Pm-3m* [23], and $CsGeBr_3$ shows a rhombohedral phase with *R3m* space group (Figure 1) [27]. Thus, we have employed cubic and rhombohedral structures for Sn and Ge-based perovskites during the initial structural optimization, respectively. To investigate the formability of the perovskite phase we have calculated the tolerance factor defined by $t = (R_A+R_X)/\sqrt{2}(R_B+R_X)$ [28], and the octahedral factor given as $\mu = R_B/R_X$ [29]($R_A$, $R_B$, and $R_X$ represent the ionic radii of $A^+$, $B^{2+}$ and $X^-$, respectively) using Shannon ionic radii. The ionic radii of $Cs^+$, $Rb^+$, $K^+$, $Na^+$, $Ge^{2+}$, $Sn^{2+}$, and $Br^-$ are given in Table S1 in the supplementary information. The calculated $t$ and $\mu$ are shown in Figure 1. The perovskite structure will exist if $t$ value is in the range of $0.813 \leq t \leq 1.107$ [30], and $\mu$ value is in the range of $0.37 \leq \mu \leq 0.859$ [4]. All the compounds except $NaSnBr_3$ have $t$ values of 0.832 – 1.009. As $NaSnBr_3$ has a $t$ value of 0.774, it does not stabilize in the perovskite phase. Hence, the $NaSnBr_3$ is not considered in further calculations. The octahedral distortion exists if the B-site cation is not likely to be well fitted with the $X_6$ octahedron, which is accountable for structural distortion. The calculated $\mu$ values are found to be in the range of $0.372 \leq \mu \leq 0.561$, as listed in Table S2. The lower values of $\mu$ for the Ge-based materials compared to the Sn-based halides indicate more octahedral distortion of Ge-based halides due to the smaller ionic radius of the $Ge^{2+}$ ions (0.73 Å) in comparison to the $Sn^{2+}$ ion (1.10 Å). This leads to more $Ge^{2+}$ cation off-centering.

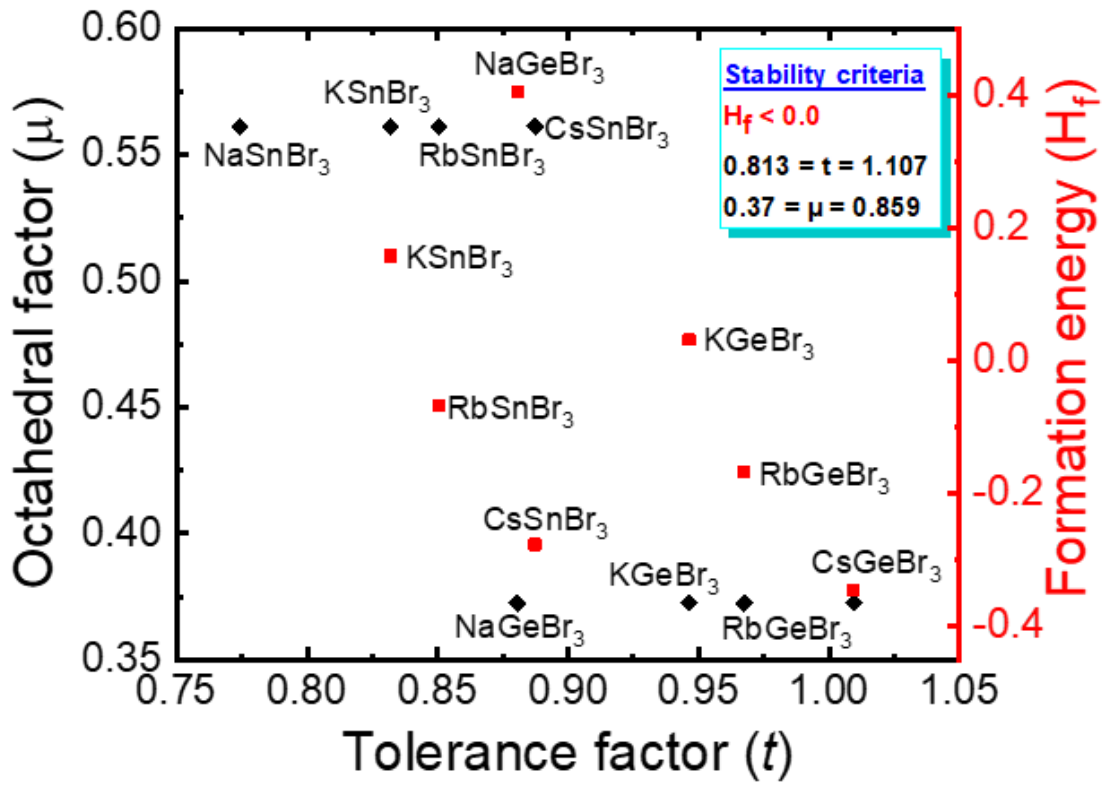

Figure 2. Calculated tolerance factor ($t$), octahedral factor ($\mu$), and formation energy of $ABBr_3$ perovskites (Black squares represent $t$ and $\mu$, while red squares represent formation energy).

The calculated lattice parameters, bond lengths, and bond angles for all the optimized structures are listed in Table 1. The lattice constants for $CsSnBr_3$ and $CsGeBr_3$ were calculated to be 5.89 Å and 5.75 Å, showed good agreement with the previously calculated results [31], [32]. Calculations slightly overestimated the lattice parameters when compared with the experiments (~1.4% for $CsSnBr_3$ and ~2.0% for $CsGeBr_3$). As the A-site cation changes from Cs to K for Sn-based, and from Cs to Na for Ge-based perovskite compounds, the lattice constant decreases slightly due to the reduction of ionic radius from Cs to Na.

Table 1. Calculated structural parameters of $ABBr_3$ perovskites using the PBE functionals.

| Compound | Lattice parameter (Å) | | | Bond length (Å) | Bond angle (deg) | | |
|---|---|---|---|---|---|---|---|
| | This work | Expt. | Others | | This work | Expt. | Others |
| $CsSnBr_3$ | 5.89 | 5.804 | 5.881 [33], 5.90 [32], 5.882 [34] | 2.945 | 90 | | |
| $RbSnBr_3$ | 5.87 | -- | 5.853 [35], 5.863 [36], 5.891[37] | 2.935 | 90 | | |
| $KSnBr_3$ | 5.85 | | 5.873 [37] | 2.925 | 90 | | |
| $CsGeBr_3$ | 5.75 | 5.635[27], [38] | 5.758 [31], 5.78 [39] | 2.5773 – 3.1875 | 89.23 | 88.74 | 88.35 [39] |
| $RbGeBr_3$ | 5.66 | -- | 5.53 [40] (PBEsol) | 2.5947 – 3.0782 | 89.40 | 87.99 [40] | |
| $KGeBr_3$ | 5.61 | | | 2.6060 – 3.0159 | 89.49 | | |
| $NaGeBr_3$ | 5.59 | | | 2.6144 – 2.9969 | 89.30 | | |

In order to investigate the thermodynamic stability of the studied compounds, the formation energies were calculated by using the formula

$$E_f = E(ABBr_3) - [E(ABr) + E(BBr_2)], \qquad (1)$$

where $E(ABBr_3)$, $E(ABr)$ and $E(BBr_2)$ represents the total energies of $ABBr_3$, *ABr*, $BBr_2$ compounds, respectively. According to eq. (1), negative value of $E_f$ indicates loss of energy during the formation of $ABBr_3$ (exothermic process), validating the thermodynamic stability of these compounds. A more negative value of $E_f$ leads to more thermodynamic stability and vice-versa. The calculated formation energies for all the compounds were depicted in Figure 1. It was observed that $CsGeBr_3$ exhibits a lower $E_f$ value compared to other materials. In contrast, $KGeBr_3$, $NaGeBr_3$ and $KSnBr_3$ showed positive $E_f$, similar to previous findings [41], indicating they are not energetically favourable. It should be noted that thermodynamic stability increases in the order of Na, K, Rb and Cs for both Ge- and Sn-halides, supporting the increase of tolerance factor from Na, K, Rb to Cs-based compounds.

**Electronic properties:** The band gap of material is an important quantity that influences the efficiency of photovoltaic materials. Semiconducting materials having suitable band gap is very crucial for the fabrication of solar cells. As majority of the Sun radiations that reaches the surface of the Earth have energy of <2 eV, photovoltaic materials with band gaps greater than 2 eV and lower than 0.9 eV are less effective. In addition, materials with a band gap range of 0.9 – 2.0 eV are effective for not only single junction SCs, but also for the top and bottom cells in the tandem structure. To probe the electronic properties we have calculated the band structures along paths connecting the high symmetry points Γ(0 0 0), X(0.5 0.0 0.0), M(0.5 0.5 0.0), R(0.5 0.5 0.5), and Γ(0 0 0) for the Sn-based; and Γ(0 0 0), Z(0.5 0.5 0.5), F(0.5 0.5 0.0), and Γ(0 0 0) for the Ge-based compounds. As shown in Figure 3, all of the studied materials exhibit direct band gaps having valence band maximum (VBM) and conduction band minimum (CBM) at R and Z points for Sn- and Ge-based perovskites, respectively. The band gap values for $CsSnBr_3$ and $CsGeBr_3$ were calculated to be 0.64 eV and 1.37 eV, respectively, in good agreement with other calculated results. However, these band gap values are quite smaller in comparison to experimental results as PBE functional results in underestimated bandgaps except from some Pb-based halide perovskites [14], [42]. As we replace the $Cs^+$ with other alkali cations, lattice constant varies in descending order from Cs to Na because of the reduction of ionic radius, resulting strengthening of p – p hybridization due to smaller distance between atoms. Consequently, VBM shifts upwards while CBM shifts downwards, resulted in the reduction of band gap. To see the effect of spin-orbit coupling (SOC) we have calculated the band gaps using PBE under SOC consideration. It was observed that under SOC consideration (PBE+SOC), band gaps for all the compounds decreases because of the splitting and downshifting of CBM [43]. Furthermore, the $\Delta_{soc}$ values for Ge-halides (0.04 eV – 0.06 eV) are very small compared to Sn-halide perovskites (0.36 eV – 0.37 eV). This indicates that Ge-based perovskites showed weak SOC effect compared to Sn-based compounds as Ge is lighter than Sn element [44]. The calculated band gap values of the studied compounds are listed in Table 2. The results are in good agreement with the previous findings [45].

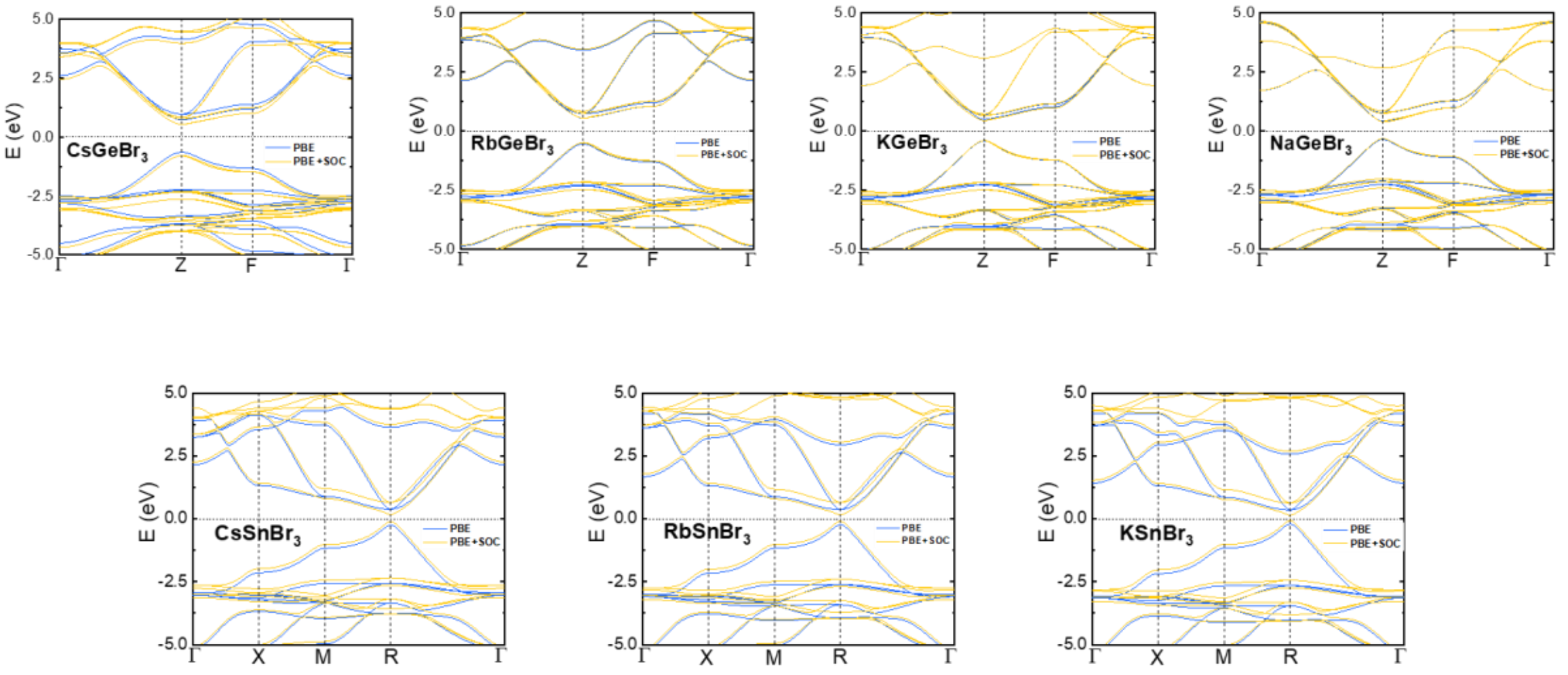

Figure 3. Band structures of studied perovskite materials (blue and yellow curves represent band structures with and without SOC using PBE functional).

Table 2. Calculated bandgap values of $ABBr_3$ perovskites studied in this work.

| Compounds | PBE | | PBE+SOC | | HSE | | Expt. |
|---|---|---|---|---|---|---|---|
| | This work | Others | This work | Others | This work | Others | |
| $CsSnBr_3$ | 0.64 | 0.64 [45], 0.63 [32], 0.615 [34] | 0.28 | | 1.10 | 1.11 [45] | 1.75 |
| $RbSnBr_3$ | 0.60 | 0.556 [35], 0.59 [36], 0.61 [37] | 0.24 | | 1.05 | 1.10 [45] | |
| $KSnBr_3$ | 0.56 | 0.57 [37] | 0.19 | | 1.00 | | |
| $CsGeBr_3$ | 1.37 | 1.44 [39], 1.53 [31] | 1.33 | 1.39 [39] | 1.97 | 1.97 [45], 1.66 [31] 2.34 [40] | 2.38 [46], 2.32 [47] |
| $RbGeBr_3$ | 1.08 | 1.57 [40] (PBEsol) | 1.03 | | 1.64 | 1.65 [45], 2.40 [40] | |
| $KGeBr_3$ | 0.91 | | 0.85 | | 1.55 | 1.47 [45] | |
| $NaGeBr_3$ | 0.78 | | 0.73 | | | | |

To get a better understanding of the electronic properties, projected density of states (PDOS) were calculated and shown in figure 4. All of the studied compounds showed similar trends of electronic orbital behaviour, the valence band is predominantly contributed by Br 4p states while the conduction band is mainly contributed by 4p/5p states of Ge/Sn elements with a little contribution from Br 4p states. In addition, below the VBM a small superposition of B-site element p states and Br p states was observed, implying a marginal hybridization of B-Br bond. It should be noted that Br 4p and Ge/Sn 4s/5s states constitutes the VBM while the CBM is formed by Br 4s and Ge/Sn 4p/5p states. Like the prototype halide perovskite $MAPbI_3$, band edges are not directly contributed by A-site elements indicating the electronically inert behaviour [48]. However, the A-site elements indirectly affect the overall electronic structure of the compound by inducing distortion of $BX_6$ octahedron due to their difference in ionic radii [49]. Moreover, to obtain further insight of electronic structure, charge densities at CBM and VBM were computed. As shown in figure 5, electron densities of VBM are located on Br and B-site element, while for CBM the densities are essentially located at B-site atom with a little presence on Br atoms. No charge accumulation around the A-site element was observed in agreement with the PDOS results.

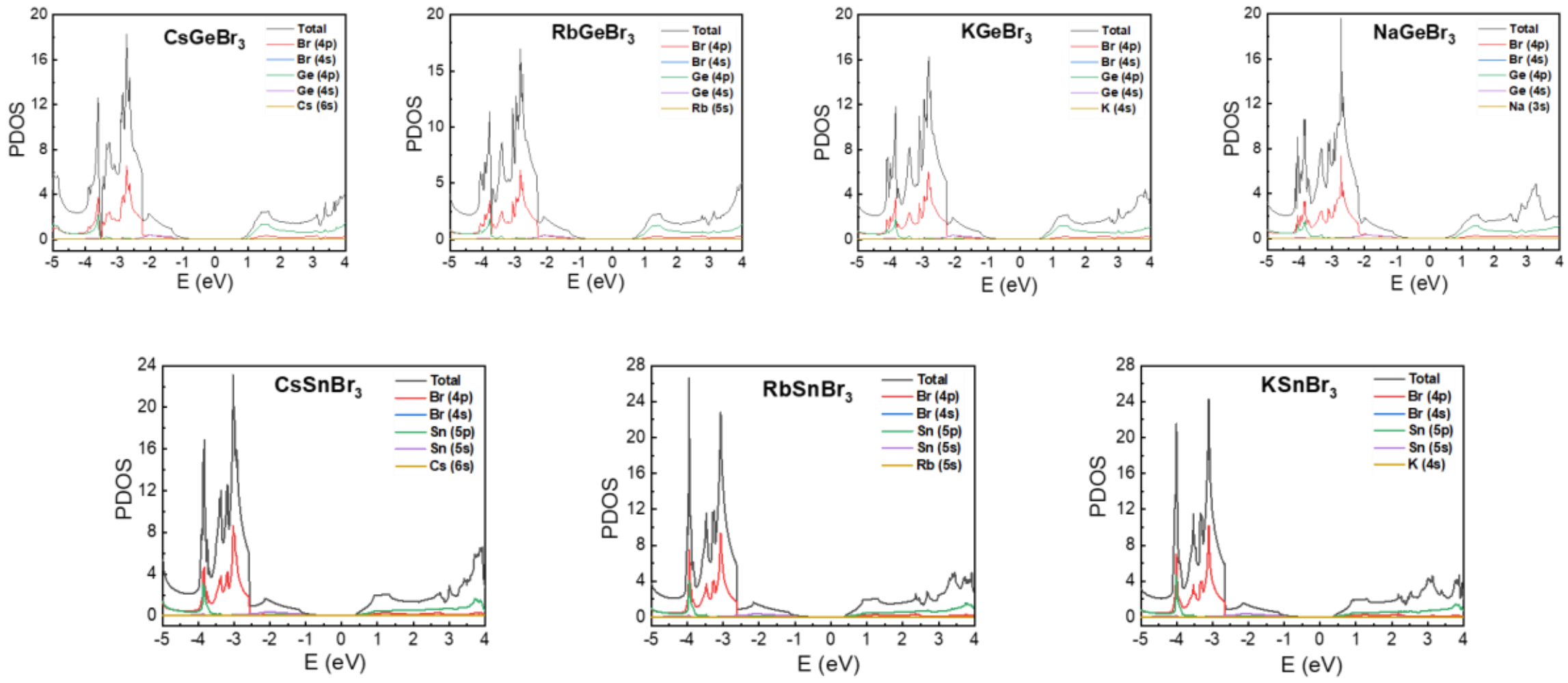

Figure 4. Calculated partial density of states, PDOS, of the studied $ABBr_3$ perovskites.

In order to analyze the charge carrier (electron and hole) transport properties of these perovskite materials, we have calculated the effective masses, which is closely related to the mobility of carriers. The lower the effective mass of carriers higher will be the carrier mobility. The effective masses of electrons and holes were calculated by fitting their energy dispersion curve to parabolic function at CBM and VBM, respectively, along two different k-paths: along Γ to Z and Z to F for Ge-perovskites; and from M to R and R to Γ for Sn-perovskites. The calculated effective masses of the carriers for all of our halide perovskites were listed in Table S3 in the supplementary information. The effective masses of holes for the Sn-based perovskites have been found to be smaller compared to Ge-based perovskite materials, attributed to the high hole mobility of Sn-halides, similar to previous theoretical findings [18], [44]. In addition, $CsGeBr_3$ showed very low effective mass for electron compared to the other investigated compounds, demonstrating very high electron mobility of $CsGeBr_3$. The change of A-site cation does not play significant role on the effective mass of carriers, the values of effective masses are more or less unchangeable.

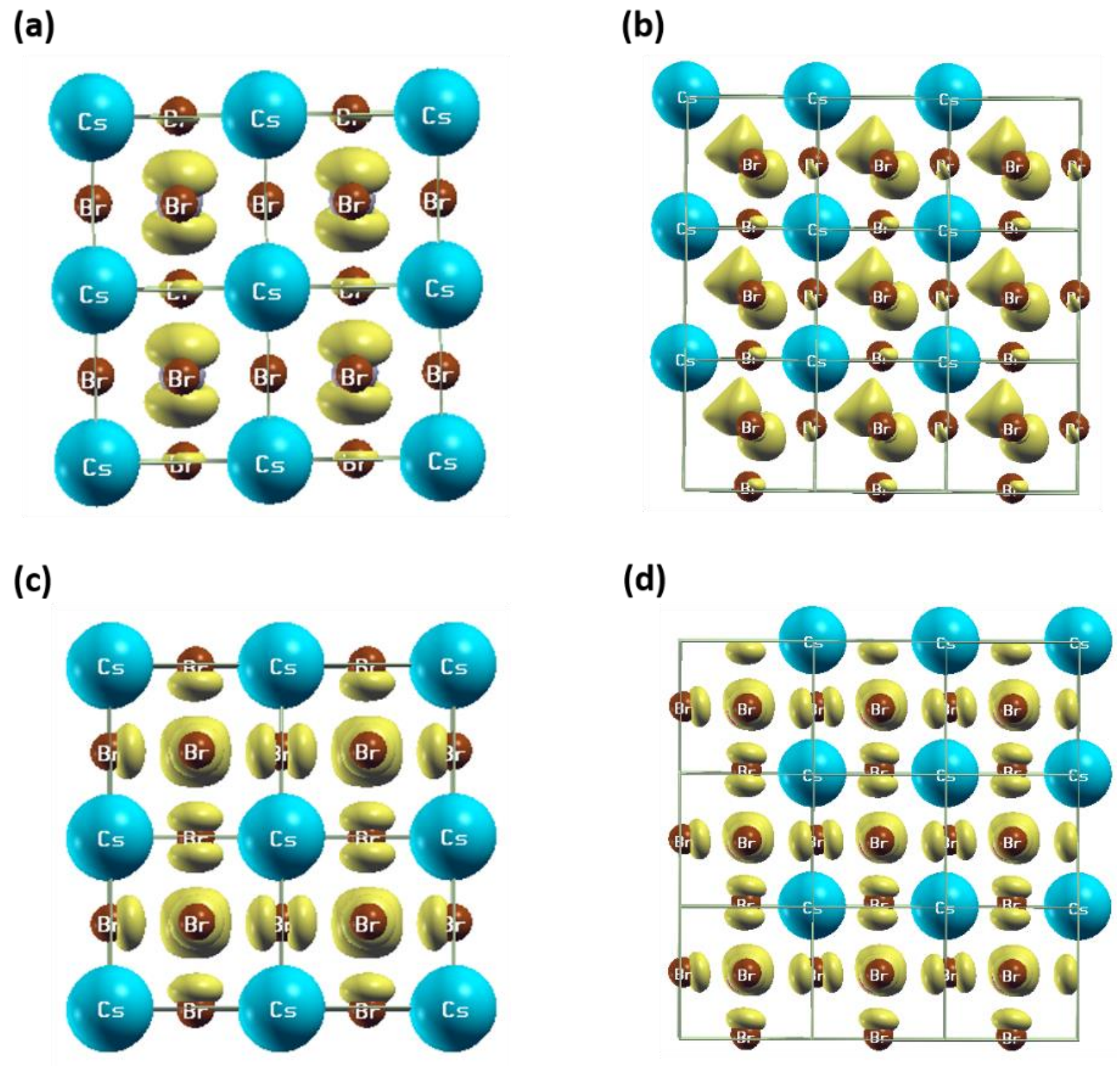

Figure 5. Charge densities of the VBM (bottom) and CBM (upper) of $CsSnBr_3$ (**a**, **c**) and $CsGeBr_3$ (**b**, **d**), respectively.

**Optical properties:** Optical properties such as absorption coefficient, dielectric constant, etc. are also very important quantities that directly impact the overall photovoltaic performance of perovskite solar cells. Large dielectric constants with high optical absorption around a wide range of solar spectrum are critical for better photovoltaic performance. The optical properties were calculated using HSE functional by the complex dielectric function $\varepsilon(\omega)$, given as

$$\varepsilon(\omega) = \varepsilon_1(\omega) + i\varepsilon_2(\omega)$$

where $\varepsilon_1(\omega)$ and $\varepsilon_2(\omega)$ represents the real and imaginary part of the dielectric constant. By using $\varepsilon_1(\omega)$ and $\varepsilon_1(\omega)$, optical properties like absorption coefficient $\alpha(\omega)$, optical conductivity $\sigma(\omega)$, reflectance $R(\omega)$, energy-loss spectrum $L(\omega)$, extinction coefficient $K(\omega)$ and refractive index $n(\omega)$ are evaluated as follows:

$$\alpha(\omega) = \sqrt{2}[\sqrt{\varepsilon_1^2(\omega) + \varepsilon_2^2(\omega)} - \varepsilon_1(\omega)]^{1/2},$$

$$\sigma(\omega) = -\frac{i\omega}{4\pi}\,\varepsilon(\omega),$$

$$R(\omega) = \left|\frac{\sqrt{\varepsilon(\omega)}-1}{\sqrt{\varepsilon(\omega)}+1}\right|^2,$$

$$L(\omega) = \frac{\varepsilon_2(\omega)}{\varepsilon_1^2(\omega)+\varepsilon_2^2(\omega)},$$

$$K(\omega) = \frac{[\sqrt{\varepsilon_1^2(\omega)+\varepsilon_2^2(\omega)} - \varepsilon_1(\omega)]^{1/2}}{2},$$

$$n(\omega) = \frac{[\sqrt{\varepsilon_1^2(\omega)+\varepsilon_2^2(\omega)}+\varepsilon_1(\omega)]^{1/2}}{2}.$$

The computed real and imaginary part of dielectric constants as a function of ω for all of our studied materials are depicted in Figure 6 The electronic part of the static dielectric constant, i.e., $\varepsilon_1(0)$, was estimated to be 4.99, 5.10, 5.18, 3.90, 4.29, and 4.46 for $CsSnBr_3$, $RbSnBr_3$, $KSnBr_3$, $CsGeBr_3$, $RbGeBr_3$, and $KGeBr_3$, respectively. The Sn-halides have larger $\varepsilon_1(0)$ values compared to Ge-halide perovskites, and upon substituting Cs with other alkali atoms $\varepsilon_1(0)$ decreases from Cs to Na. The principal peak values of $\varepsilon_1(\omega)$ for $CsSnBr_3$, $RbSnBr_3$, $KSnBr_3$, $CsGeBr_3$, $RbGeBr_3$, and $KGeBr_3$ were found to be 5.81, 6.00, 5.95, 5.00, 5.42 and 5.29 at 1.01, 1.00, 0.83, 2.25, 2.05 and 1.54 eV, respectively. The imaginary part of dielectric constant $\varepsilon_2(\omega)$ is related to the density of states of the material and describes the absorption characteristics [50]. It was observed that $\varepsilon_2(\omega)$ is red-shifted with the change of A-site cation from Cs to Na for both Sn and Ge-based materials and it can be ascribed to the corresponding band gap reduction. The critical onset points in the $\varepsilon_2(\omega)$ were observed at 1.01, 1.00, 0.99, 1.85, 1.54 and 1.37 eV for $CsSnBr_3$, $RbSnBr_3$, $KSnBr_3$, $CsGeBr_3$, $RbGeBr_3$, and $KGeBr_3$, respectively, related to the corresponding calculated band gap values. Similar characteristics were observed for the K(ω) spectra as shown in Figure S3.

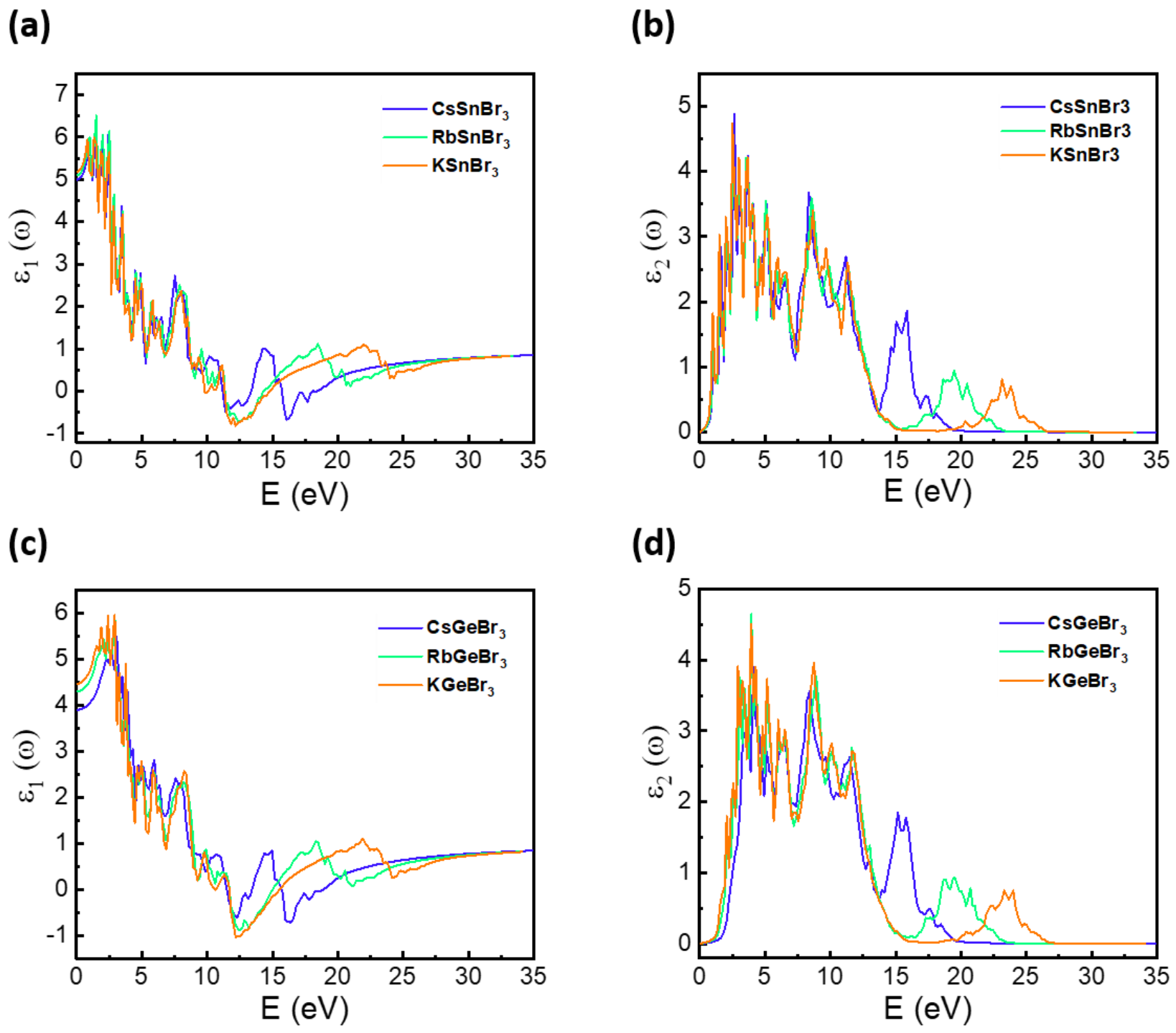

Figure 6. Spectra of dielectric constants: (a, c) real and (b, d) imaginary part of the investigated materials.

In addition, other optical properties like energy-loss L(ω), refractive index n(ω), and reflectance R(ω) also impact the performance of solar cell. The L(ω) describes the energy loss of electrons when they are propagated through the system, and the peaks in the L(ω) spectra (Figure S 3) represents the plasma resonance [42], [51]. The most prominent peaks were found to be at 14.26, 18.71, 13.80, 15.01, 19.07 and 15.60 eV for $CsSnBr_3$, $RbSnBr_3$, $KSnBr_3$, $CsGeBr_3$, $RbGeBr_3$, and $KGeBr_3$, respectively. The refractive index is also an essential parameter of materials that represents the amount of light refracted, which is related to the microscopic atomic interactions [52]. The computed n(ω) spectra were depicted in Figure S 4. The static refractive index n(0) values were calculated to be 2.23, 2.26, 2.28, 1.97, 2.07 and 2.11 for $CsSnBr_3$, $RbSnBr_3$, $KSnBr_3$, $CsGeBr_3$, $RbGeBr_3$, and $KGeBr_3$, respectively. It should be noted that with the change of alkali elements, the static refractive index increases from Cs to K, and Cs to Na for Sn-and Ge-based materials, respectively. As the energy increases, n(ω) increases and reaching a maximum value, and then decreases gradually and goes below unity for certain energy ranges for all the studied compounds. At these energy ranges group velocity of the incident radiation surpasses the velocity of light.

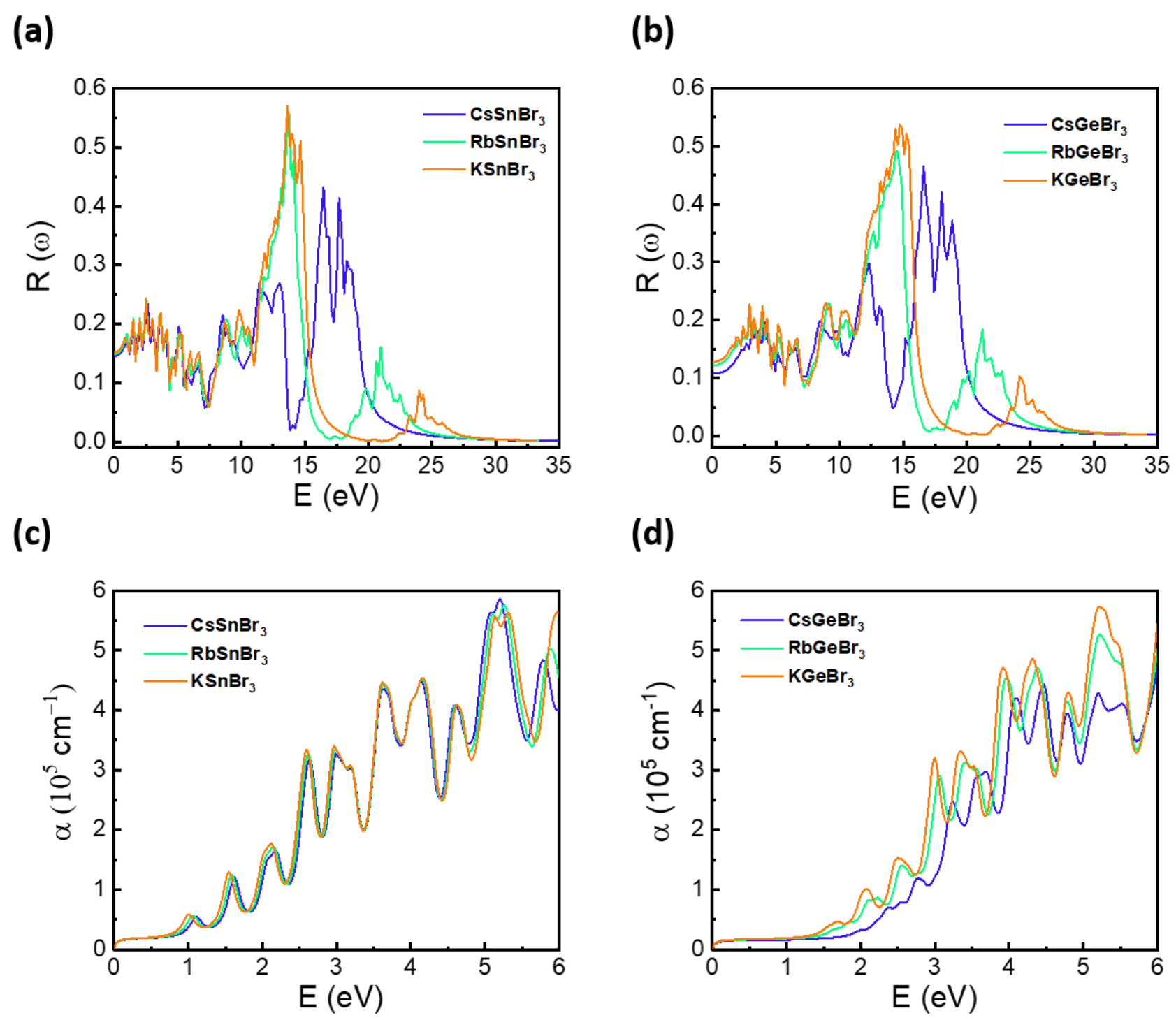

Figure 7. Reflectance (**a, b**) and absorbance (**c, d**) spectra of the $ABBr_3$ perovskites.

The reflectivity R(ω) of a material determines amount of incident radiation that would be reflected from it. As shown in Figure 7, in the low-energy range (<4 eV), reflectivity reaches 24.3% and 22.7% for Sn- and Ge-based perovskites, respectively, demonstrating good transparency in the visible and UV regions. Furthermore, for both Sn and Ge halides, the static reflectivity R(0) decreases as the atomic radius of A-site element decreases. The absorption coefficient is another crucial parameter that significantly influences the performance of solar cells. It is noted that for both Ge- and Sn-based perovskite materials the absorption onset and the first peak of absorption are red shifted as the atomic number of A-site element decreases. These findings remarkably consistent with the predicted band structures and DOS. All the studied materials exhibited strong absorption in the visible and UV range, and is higher than that of the prototype perovskite material $MAPbI_3$ ($3.8 \times 10^4$ $cm^{-1}$) [53], [54]. In terms of A-site element, K-based perovskites exhibited maximum absorption compared to Rb and Cs-based materials for both Ge and Sn halide perovskites. Moreover, all the studied materials showed moderate absorption in the infrared region. The desirable absorption properties of all the investigated materials render them as potential candidates for absorber material of solar cell.

**Conclusion:** In summary, employing DFT-based first principles calculation, we have studied the structural, electronic, and optical properties of eco-friendly $ABBr_3$ perovskites. The findings demonstrates that both Cs and Rb based perovskites are energetically favourable. All the studied compounds exhibit direct bandgap ranging from 1.00 eV to 1.97 eV, desirable for solar cell application. Furthermore, all the $ABBr_3$ perovskites exhibit strong optical absorption in the entire UV-vis range. We believe that this work will inspire researchers to experimentally investigate these materials as well as stimulate further research into alternative lead-free perovskite materials.

# Supplementary Information

Dibyajyoti Saikia[1], Mahfooz Alam[2], Jayanta Bera[1], Atanu Betal[1], Appala Naidu Gandhi[2], Satyajit Sahu[1]

[1]Department of Physics, Indian Institute of Technology Jodhpur, Jodhpur, India 342037

[2]Department of Metallurgical and Materials Engineering, Indian Institute of Technology Jodhpur, Jodhpur, India 342037

Table S1: The ionic radius of $ABBr_3$ compounds studied in this work.

| A-cations | Radius (Å) | B-cations | Radius (Å) | X-cations | Radius (Å) |
|---|---|---|---|---|---|
| Cs | 1.88 | Sn | 0.73 | Br | 1.96 |
| Rb | 1.72 | Ge | 1.10 | | |
| K | 1.64 | | | | |
| Na | 1.39 | | | | |

Table S2: Selected compounds and corresponding space groups used for formation energy calculation.

| Compounds | Space group |
|---|---|
| CsBr | Fm-3m |
| RbBr | Fm-3m |
| KBr | Fm-3m |
| NaBr | Fm-3m |
| $SnBr_2$ | $P4_2/mnm$ |
| $GeBr_2$ | $P12_1/C1$ |

Table S3: Calculated tolerance factor (*t*), octahedral factor (*μ*) and formation energy ($E_f$) of $ABBr_3$ compounds studied in this work.

| Parameters | $CsGeBr_3$ | $RbGeBr_3$ | $KGeBr_3$ | $NaGeBr_3$ | $CsSnBr_3$ | $RbSnBr_3$ | $KSnBr_3$ | $NaSnBr_3$ |
|---|---|---|---|---|---|---|---|---|
| *t* | 1.009 | 0.967 | 0.946 | 0.881 | 0.887 | 0.850 | 0.832 | 0.774 |
| *μ* | 0.372 | 0.372 | 0.372 | 0.372 | 0.561 | 0.561 | 0.561 | 0.561 |
| $E_f$ (eV) | -0.345 | -0.167 | 0.031 | 0.405 | -0.277 | -0.067 | 0.158 | -- |

Table S3. Calculated electron and hole effective masses of the $ABBr_3$ compounds studied in this work.

| Compounds | Direction | $m_e^*/m_0$ | $m_{\mathrm{h}}^*/m_0$ | $m_r^*/m_0$ |
|---|---|---|---|---|
| $CsSnBr_3$ | R to Γ | 0.28 | 0.11 | 0.07 |
| | M to R | 0.90 | 0.12 | 0.10 |
| $RbSnBr_3$ | R to Γ | 0.27 | 0.11 | 0.07 |
| | M to R | 0.88 | 0.12 | 0.10 |
| $KSnBr_3$ | R to Γ | 0.27 | 0.14 | 0.09 |
| | M to R | 0.88 | 0.18 | 0.15 |
| $CsGeBr_3$ | Γ to Z | 0.02 | 0.22 | 0.02 |
| | Z to F | 0.09 | 0.30 | 0.07 |
| $RbGeBr_3$ | Γ to Z | 0.16 | 0.16 | 0.08 |
| | Z to F | 1.2 | 0.23 | 0.19 |
| $KGeBr_3$ | Γ to Z | 0.14 | 0.16 | 0.07 |
| | Z to F | 0.87 | 0.21 | 0.17 |
| $NaGeBr_3$ | Γ to Z | 0.13 | 0.16 | 0.07 |
| | Z to F | 0.84 | 0.26 | 0.20 |

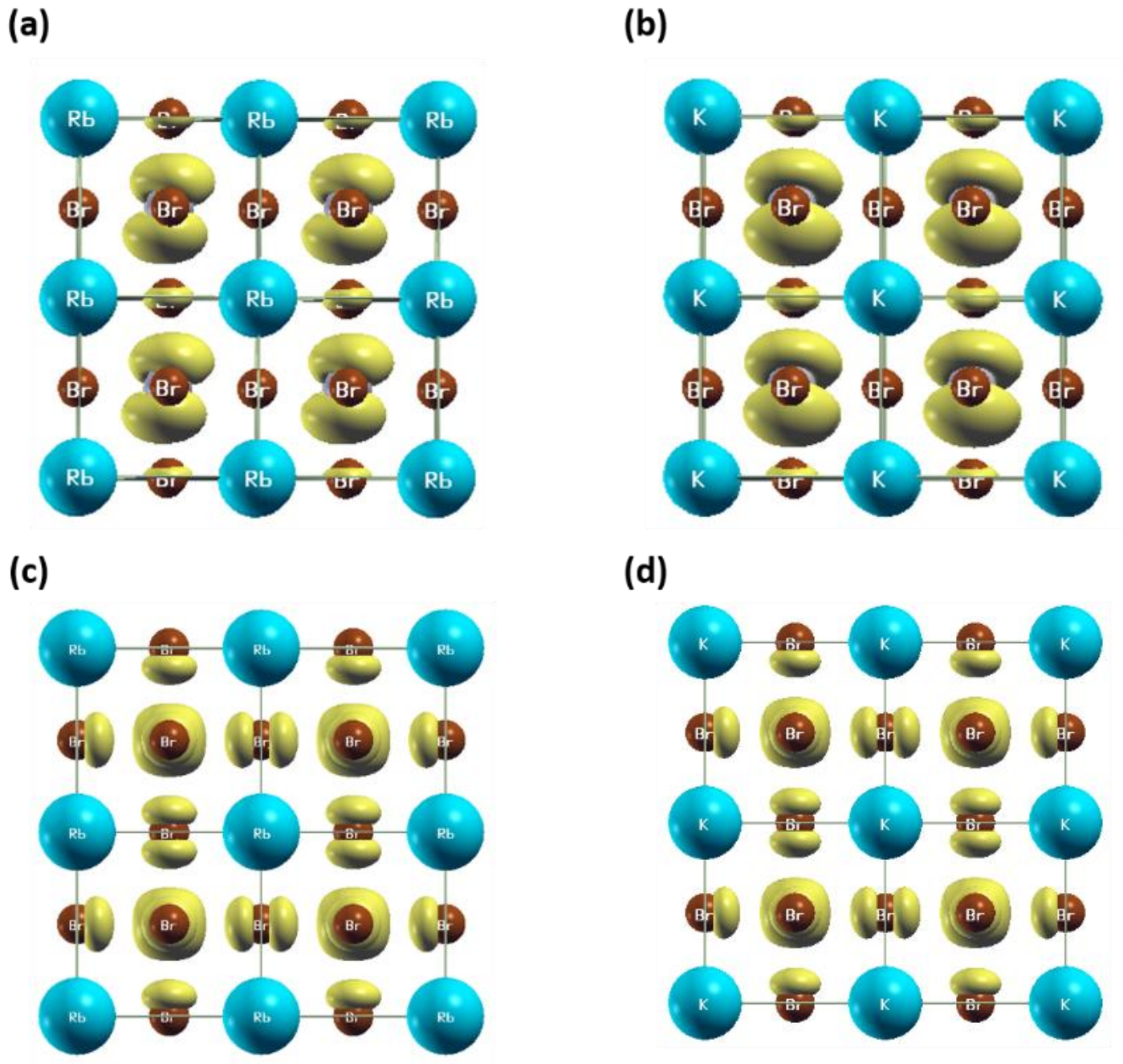

Figure S1. Charge densities of the VBM (bottom) and CBM (upper) of $RbSnBr_3$ (**a**, **c**) and $KSnBr_3$ (**b**, **d**), respectively.

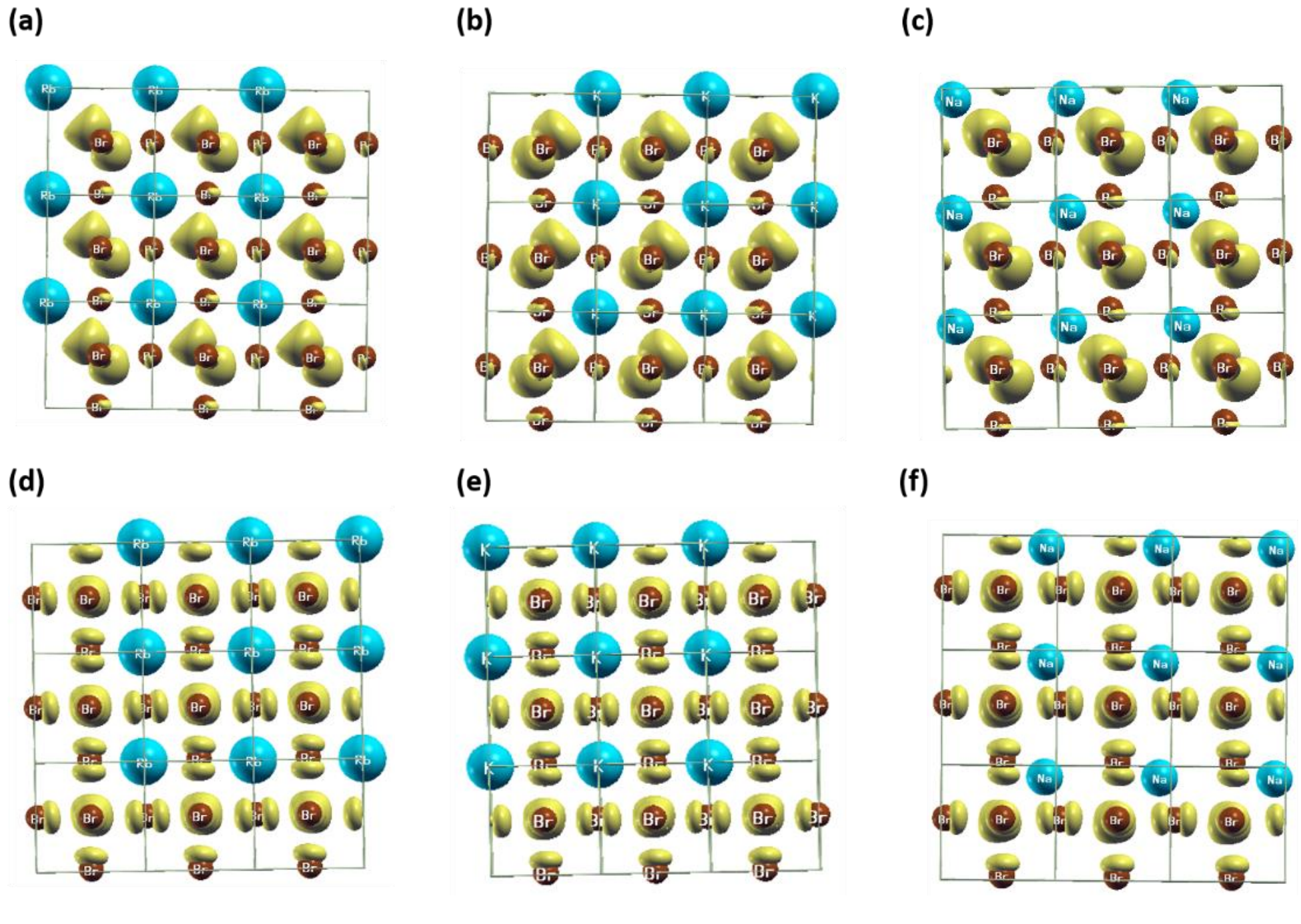

Figure S2. Charge densities of the VBM (bottom) and CBM (upper) of $RbGeBr_3$ (**a**, **d**) $KGeBr_3$ (**b**, **e**) and $NaGeBr_3$ (**c, f**), respectively.

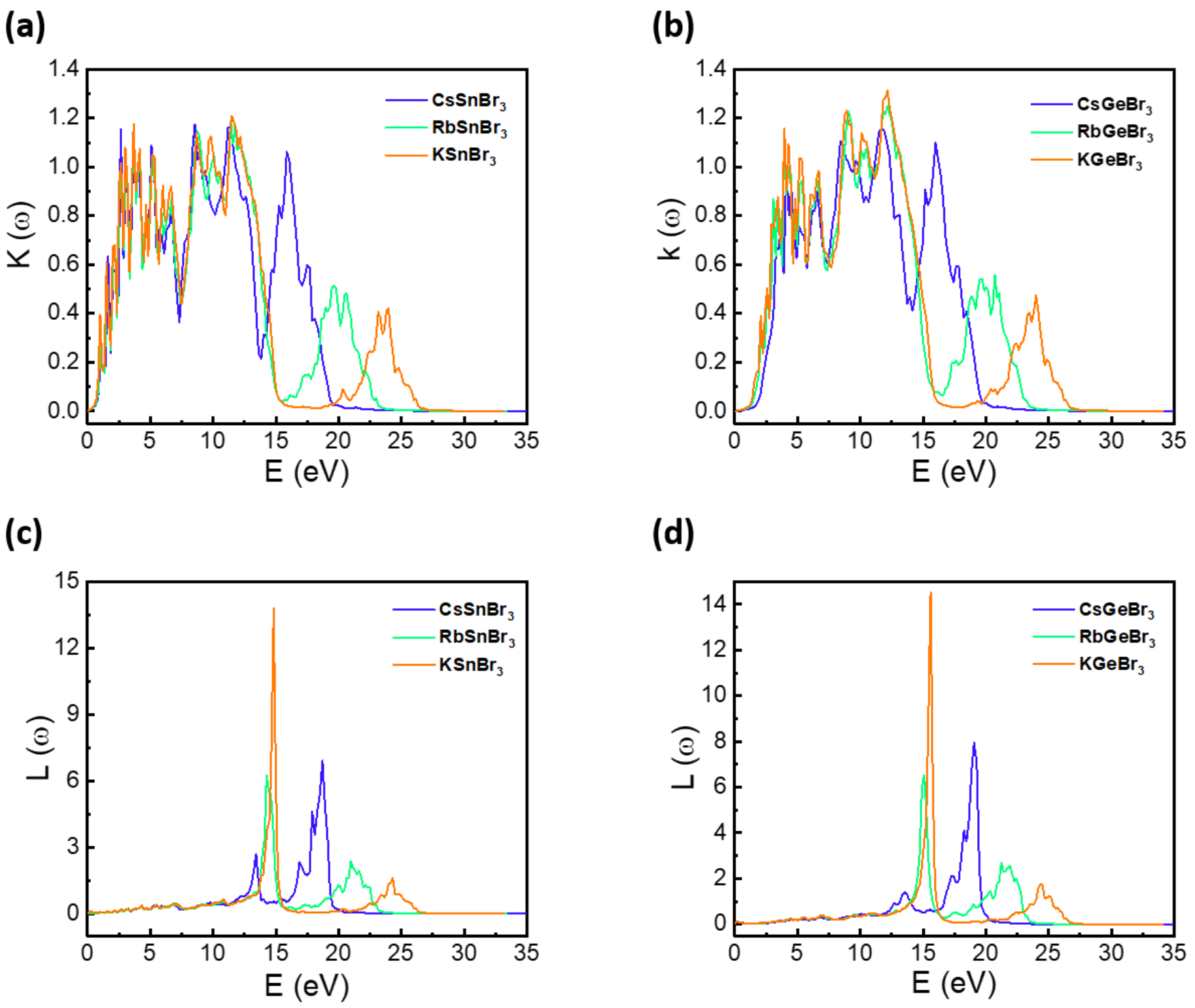

Figure S3. Spectra of extinction coefficients (**a**, **c**) and electron energy loss (**b**, **d**) of the studied $ABBr_3$ materials in this work.

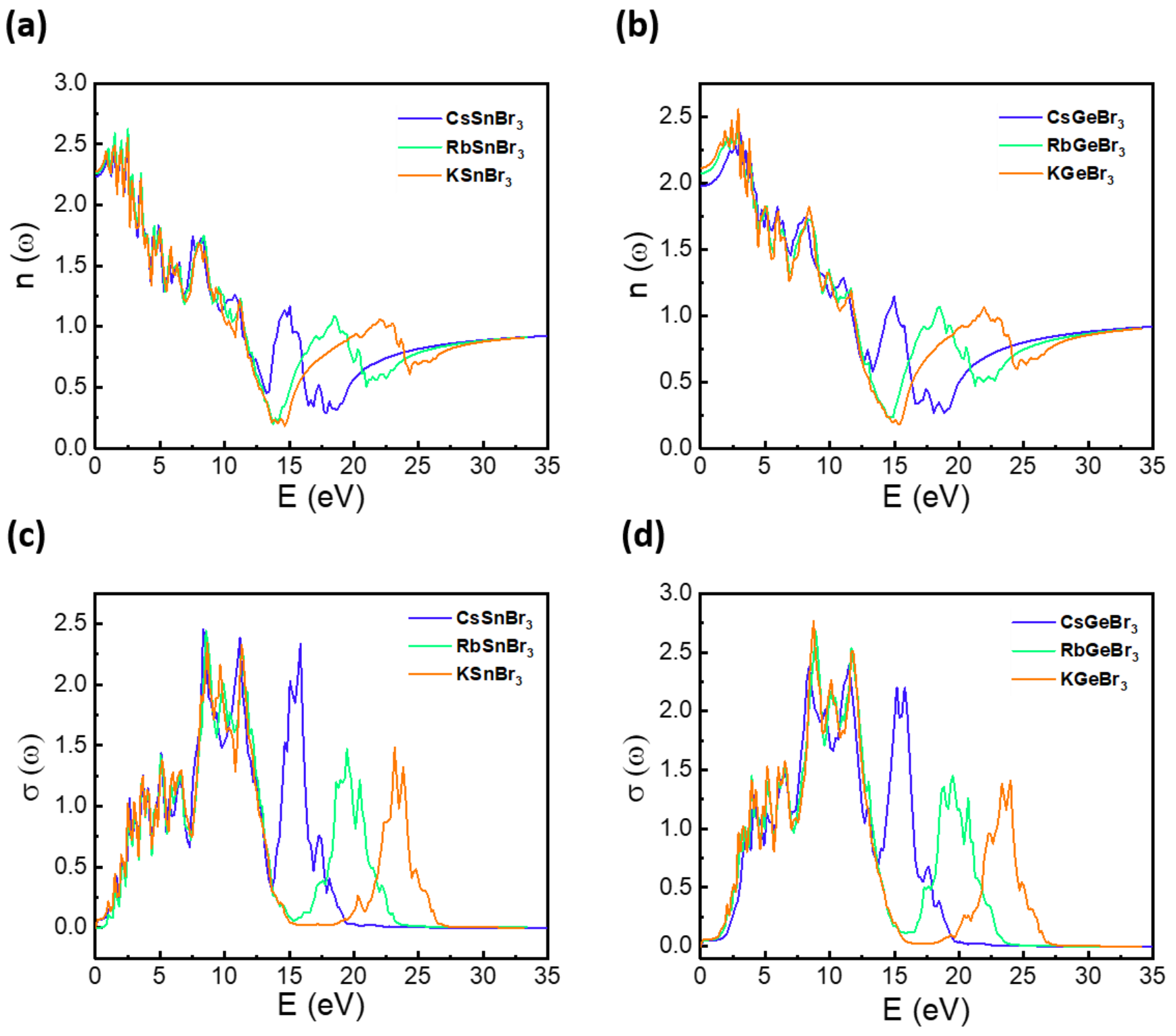

Figure S4. Spectra of refractive index (**a**, **b**) and conductivity (**c**, **d**) of the studied $ABBr_3$ materials in this work.